\documentclass[10pt,times]{iopart}
\usepackage{epsfig}
\usepackage{graphics}
\usepackage{graphicx}
\usepackage{bm}
\usepackage{iopams}
\usepackage{color}

\begin{document}

\noindent AEI-2006-030


\newcommand{\geo}{\textsc{GEO\,600}}
\newcommand{\Ex}{{\bf E}_{\rm X}}
\newcommand{\dEx}{\Delta {\bf E}_{\rm X}}
\newcommand{\dEh}{\Delta {\bf E}_{\rm H}}
\newcommand{\dE}{\Delta {\bf E}}
\newcommand{\Exa}{E^{a}_{\rm X}}
\newcommand{\Exf}{E^{f}_{\rm X}}
\newcommand{\Ext}{E^{t}_{\rm X}}
\newcommand{\Exb}{E^{b}_{\rm X}}
\newcommand{\Eh}{{\bf E}_{\rm H}}
\newcommand{\Eha}{E^{a}_{\rm H}}
\newcommand{\Ehf}{E^{f}_{\rm H}}
\newcommand{\Eht}{E^{t}_{\rm H}}
\newcommand{\x}{$X$}
\newcommand{\h}{$H$}
\newcommand{\df}{\mathrm{d}f}
\newcommand{\dt}{\mathrm{d}t}
\newcommand{\dx}{\mathrm{d}x}
\newcommand{\dwf}{\mathrm{d}w^f}
\newcommand{\dwt}{\mathrm{d}w^t}
\newcommand{\dwa}{\mathrm{d}w^a}
\newcommand{\Txhf}{T_\mathrm{XH}(f)}
\newcommand{\Sx}{{\bf S}_{\rm X}}
\newcommand{\Sh}{{\bf S}_{\rm H}}
\newcommand{\bH}{\bf \tilde H}
\newcommand{\bX}{\bf \tilde X}
\newcommand{\btdelta}{\bm{\tilde \delta}}

\title[Instrumental Vetoes]{Robust vetoes for gravitational-wave 
burst triggers using known instrumental couplings}

\author{P~Ajith\dag, M~Hewitson\dag, J~R~Smith\dag~and K~A~Strain\dag\ddag}

\address{\dag~Max-Planck-Institut f\"ur Gravitationsphysik (Albert-Einstein-Institut) \\
und Universit\"at Hannover, Callinstr.~38, 30167 Hannover, Germany}
\address{\ddag~Department of Physics and Astronomy, University of Glasgow, \\
Glasgow, G12 8QQ, United Kingdom}

\eads{\mailto{Ajith.Parameswaran@aei.mpg.de}}

\date{\today}


\begin{abstract}

The search for signatures of transient, unmodelled gravitational-wave (GW) bursts in the 
data of ground-based interferometric detectors typically uses `excess-power' search methods. 
One of the most challenging problems in the burst-data-analysis is to distinguish between 
actual GW bursts and spurious noise transients that trigger the detection algorithms. In 
this paper, we present a unique and robust strategy to `veto' the instrumental glitches. 
This method makes use of the phenomenological understanding of the coupling of different 
detector sub-systems to the main detector output. The main idea behind this method is that
the noise at the detector output (channel \h) can be projected into two orthogonal directions 
in the Fourier space -- along, and orthogonal to, the direction in which the noise in 
an instrumental channel \x would couple into \h. If a noise transient in the detector 
output originates from channel \x, it leaves the statistics of the noise-component of \h\ 
orthogonal to \x\ unchanged, which can be verified by a statistical hypothesis testing. This 
strategy is demonstrated by doing software injections in simulated Gaussian noise.  We also 
formulate a less-rigorous, but computationally inexpensive alternative to the above method. 
Here, the parameters of the triggers in channel \x\ are compared to the parameters of the 
triggers in channel \h\ to see whether a trigger in channel \h\ can be `explained' by a 
trigger in channel \x\ and the measured transfer function. 

\end{abstract}



\section {Introduction}

A world-wide network of gravitational-wave (GW) detectors consisting of ground-based
interferometers \cite{ligo,virgo,geo,tama} and resonant-bars \cite{explNautilus,auriga,allegro} 
has now started looking for signatures of GWs expected to be coming from astrophysical and 
cosmological sources. Among the most promising astrophysical sources of GWs for these 
ground-based detectors are the transient, unmodelled astrophysical phenomena like core-collapse 
supernovae, Gamma-ray bursts and black hole/neutron star mergers. Although, in recent 
years, numerical relativity has made a tremendous progress in predicting the gravitational
waveforms from compact binary mergers and gravitational collapses (see, for e.g., \cite{numRel} 
for some recent work), these predictions are not yet robust enough so as to allow the 
data-analysts to use techniques like matched filtering, which relies on the accurate models 
of waveforms. Most of the algorithms currently used in burst-searches are time-frequency 
detection algorithms which look for short-lived excitations of power in time-frequency maps 
constructed from the time-series data \cite{ExPower,TFClusters,Waveburst}. Such detection 
algorithms implemented in the data analysis pipelines are usually called event trigger 
generators (ETGs). 

Since current interferometric detectors are highly complex instruments, the detector
output will contain a large number of spurious noise transients which trigger the ETGs. 
One of the main challenges in the burst-data-analysis is to distinguish these 
instrumental `glitches' from actual GW bursts. One way of approaching the problem is to 
use more than one detector (mostly located at different parts of the world) and look for 
coincident burst signals in these data streams. While coincident instrumental bursts in 
multiple detectors are highly improbable, long data-taking runs (typically several 
months long) with multiple detectors can produce a large number of random coincidences. 
All of the detector groups are thus developing various techniques to `veto' the spurious 
instrumental bursts at the detector site itself in order to reduce the list of candidate 
coincident triggers (see \cite{PQMon,DiCredico05,hNull} for some recent work in this direction).  
   
In this paper we present a robust veto strategy which makes use of a phenomenological 
understanding of the coupling of various detector sub-systems with the main detector output. 
This provides us the possibility of vetoing the triggers with a very high confidence. Although 
this method was developed for the \geo~\cite{geo} detector, this can be applied to the
data from any interferometric GW detector. 

The main idea behind this method is that the noise measured at the detector output
can be decomposed into two orthogonal components. If the physical coupling between a 
sub-system $X$ and the detector output is measured, this information can be used to 
{\it transfer}, or {\it map}, the noise in $X$ to the detector output, and hence the 
component of $H$ orthogonal to the `mapped' $X$ can also be calculated. If a noise transient
originating from the sub-system $X$ appears in the detector output, it leaves the statistics
of the noise in the direction orthogonal to the mapped $X$ unchanged. In Sections \ref{sec:NoiseProjVeto} 
and \ref{sec:NPVetoImpl} we formulate and demonstrate a strategy to veto such noise transients using 
the \emph{transfer function} from $X$ to $H$. In Section \ref{sec:EvntProjVeto}, we develop an alternative veto 
strategy using the parameters of the burst triggers estimated by the ETG.


\section {Vetoes using known instrumental couplings}
\label{sec:NoiseProjVeto}

Let $x(t)$ and $h(t)$ denote the time-series data measured at the input and output of 
a linear, time-invariant system. The input and output of the system are related by
the {\it transfer function} $T_\mathrm{XH}(s)$ of the system, defined as \cite{dsp}
\begin{equation}
T_\mathrm{XH}(s) = \frac{H(s)}{X(s)},
\end{equation}
where $H(s)$ and $X(s)$ represent the Laplace transforms of $x(t)$ and $h(t)$, respectively,
and the complex variable $s$ represents a point in the Laplace space. Although the transfer 
function $T_\mathrm{XH}(s)$ is formally defined in the Laplace space, for the purpose of this 
paper, it is easier to work in the Fourier domain. If $\tilde X(f)$ and $\tilde H(f)$ are the 
Fourier transforms of $x(t)$ and $h(t)$, respectively, the equivalent relation in the Fourier
domain is given by   
\begin{equation}
T_\mathrm{XH}(f) = \frac{\tilde H(f)}{\tilde X(f)}\,,
\end{equation}
or, 
\begin{equation}
T_\mathrm{XH}(f) = \frac{P_\mathrm{XH}}{P_\mathrm{XX}}\,,
\end{equation}
where $P_\mathrm{XH}$ is the cross-power spectral density of $x(t)$ and $h(t)$, and $P_\mathrm{XX}$
is the power spectral density of $x(t)$. In the formal language of signal processing, 
$T_\mathrm{XH}(f)$ is described by the {\it frequency response} (magnitude of $T_\mathrm{XH}(f)$) 
and {\it phase-shift} (phase of $T_\mathrm{XH}(f)$). But we will conveniently think of 
$T_\mathrm{XH}(f)$ and $T_\mathrm{XH}(s)$ as two different representations of the transfer 
function. 

We will refer to the measurement points for time-series data within the detector as `channels' 
and assume that they are continuously recorded. Let $x(t)$ and $h(t)$ denote the time-series 
data in channel \x\ (which, presumably records the noise from a detector sub-system) and channel 
\h\ (the main detector output), respectively. The transfer function from \x\ to \h\ can be 
measured by injecting some noise in \x\ and measuring $\tilde X(f)$ and $\tilde H(f)$ 
simultaneously~\cite{NoiseProj}. This is done in such a way that the injected noise from \x\ 
completely dominates channel \h, and the contributions from other noise sources are negligible. 
The measured transfer function represents our phenomenological understanding of the physical 
coupling of a detector sub-system to the main detector output. If we assume that the coupling 
of noise between channel \x\ and \h\ is linear and the transfer function is time-invariant, 
the Fourier transform of noise measured in channel \x\ at any time can be {\it transferred} 
to channel \h, by
\begin{equation}
\tilde X'(f) =  T_\mathrm{XH}(f) \, \tilde X(f).
\label{eq:noiseProj}
\end{equation}

$\tilde X(f)$ and $\tilde H(f)$ can be thought of as components of the vectors $\bX$ 
and $\bH$, defined in two different infinite-dimensional Hilbert spaces spanned 
by the Fourier basis functions. In that sense, Eq.(\ref{eq:noiseProj}) is equivalent to 
mapping $\bX$ into the space of $\bH$. Then, the component of $\bH$ that is orthogonal to 
$\bX'$ can be found by a Gram-Schmidt orthogonalization~\cite{Arfken}
\begin{equation}
\btdelta = \bH - \mathrm{proj}\,_{\bX'} \bH \, ,
\label{eq:delta}
\end{equation}
where we define the projection operator by
\begin{equation}
\mathrm{proj}\,_{\bf \tilde u}{\bf \tilde v}= \frac{\left<{\bf \tilde v},{\bf \tilde u}\right>}
{\left<{\bf \tilde u},{\bf \tilde u}\right>}\,{\bf \tilde u} \,.
\end{equation}
In the above expression $\left<{\bf \tilde v},{\bf \tilde u}\right>$ denotes the inner 
product between the vectors $\bf \tilde v$ and $\bf \tilde u$: 
\begin{equation}
\left<{\bf \tilde v},{\bf \tilde u}\right> = 2\int_0^\infty \tilde v(f) \, \tilde u(f)^* \, \df.
\end{equation}

All of the continuous quantities like $\tilde H(f)$ and  $\tilde X(f)$ can be recast 
in terms of their discrete counterparts like $\tilde H_k$ and  $\tilde X_k$, where the 
index $k$ represents the $k$th frequency bin of the discrete Fourier transform (DFT). 
Now $\bf \tilde H$ and $\bf \tilde X$ can be thought of as vectors defined in two $N$-dimensional
Hilbert spaces, where $N$ is the number of data samples used to compute the DFTs. 

The noise in different channels is generated by different random processes and 
has specific statistical and spectral properties. If a non-stationarity (e.g.,~a glitch)
originates in channel \x, it will change the statistical properties of that 
segment of data in channel \x, and hence, in channel \h. But the statistical 
properties of $\bm{\tilde \delta}$ constructed from this segment will remain unaffected.
On the other hand, if the non-stationarity does {\it not} originate in \x, it {\it will} change
the statistical properties of $\bm{\tilde \delta}$ constructed from this segment. 
This can be verified by a statistical hypothesis testing. If $\bm{\tilde \delta}$ 
constructed from the segment of data containing the non-stationarity is statistically 
similar to that constructed from the neighboring segments, this means
that the non-stationarity originates in an instrumental sub-system (channel \x)
and we veto the trigger. If this is not true, this means that the non-stationarity does 
{\it not} originate in channel \x, and we keep the trigger. This exercise can be 
repeated with all the known noise sources (all the measured channels that are known to 
couple to the detector output). 

\subsection{Test statistic}
\label{sec:TestStat}

If the components of $\bH$ are generated by zero-mean Gaussian processes, the real and imaginary 
parts of $\tilde \delta_k$ in each frequency bin will be distributed according to a Gaussian 
distribution of mean zero and variance $\sigma_k^2$. Following~\cite{ExPower}, we compute 
the `excess-power' statistic from $\btdelta$: 
\begin{equation}
\epsilon = \sum_{k=m}^{m+M} P_k \,,~~\, P_k = \frac{|\tilde \delta_k|^2}{\sigma_k^2}. 
\label{eq:stat}
\end{equation}
It can be shown that $\epsilon$ will follow a $\chi^2$ distribution of $2M$ degrees of 
freedom in the case of a non-windowed DFT. But in the case of a windowed DFT, $P_k$ are not 
{\it independent} $\chi^2$ variables, and hence $\epsilon$ will {\it not} follow a $\chi^2$ 
distribution~\cite{JL}. But to a very good approximation, $\epsilon$ will follow a Gamma
distribution with scale parameter $\alpha$ and shape parameter $\beta$. These parameters
are related to the mean and variance of the distribution of $\epsilon$ by
\begin{equation}
\alpha = \Big(\frac{\mu_\epsilon}{\sigma_\epsilon}\Big)^2, \,\,\,\, 
\beta = \frac{\sigma^2_\epsilon}{\mu_\epsilon}.
\end{equation} 
In order to estimate the parameters of the Gamma distribution, we generate a population of 
$\epsilon$ from stationary data surrounding the burst (using the same DFT-length and bandwidth). 
From that population, $\mu_\epsilon$ and $\sigma^2_\epsilon$ can be estimated, and hence 
$\alpha$ and $\beta$. 

If the computed $\epsilon$ (from the segment of data containing the burst) is less than a 
threshold, we veto the trigger. The threshold $\tau$ giving a rejection probability of $\psi$ 
can be found by 
\begin{equation}
\psi = \int_0^\tau \Gamma(x;\alpha,\beta) \, \dx,
\label{eq:thresh}
\end{equation}
where $\Gamma(x;\alpha,\beta)$ is the probability density of the Gamma distribution with 
parameters $\alpha$ and $\beta$.

\section{Implementation}
\label{sec:NPVetoImpl}
Two sets of burst triggers are generated by running an ETG on channels \x\ and \h.  
We take a set of triggers that are coincident in channels \x\ and \h, allowing a 
liberal time-window for coincidence. The data is divided into $N$ number of segments,
each of length $L$. The test statistic, $\epsilon$, is computed from the segment of 
data containing the burst. It is well-known that the maximum signal-to-noise ratio 
(SNR) for the `excess power' statistic is achieved when the time-frequency volume 
used to compute the statistic is equal to the actual time-frequency volume of the 
signal~\cite{ExPower}. Since the duration and bandwidth of the burst is estimated 
by the ETG itself, this information is used to decide on the length ($L$) of the 
data-segment used to compute $\tilde \delta_k$ and the bandwidth over which the 
integration is carried out in Eq.(\ref{eq:stat}). Consequently, the frequency resolution 
of the DFT used in the analysis is in general different for each trigger, and hence, 
so are the dimension of the vectors $\bf \tilde H$ and $\bf \tilde X$. It is then 
required that the discrete transfer function vector should also have the same dimension. 
So we store a high-resolution transfer function and interpolate it to the required 
frequency resolution. It was found that the analysis can be sensitive to the errors 
in the interpolation, since the interpolation can smear out the detailed features in the 
transfer function. Since, the lower the frequency resolution the higher are the errors, 
we set up a minimum frequency resolution of 16 Hz for the analysis. In order to achieve 
this, the minimum length of the data segment used to compute the DFT is set to be 1/16 
s $\simeq$ 60 ms. 

The parameters of the Gamma distribution are estimated from segments of data neighboring the one 
containing the burst, but excluding that segment. The trigger is vetoed if 
$\epsilon \leq \tau$, where the threshold $\tau$ giving a particular rejection probability
$\psi$ is calculated using Eq.(\ref{eq:thresh}). 

\subsection{Software injections}
\label{sec:SWinj}

\begin{figure}[tb]
\centering
\includegraphics[width=4.1in]{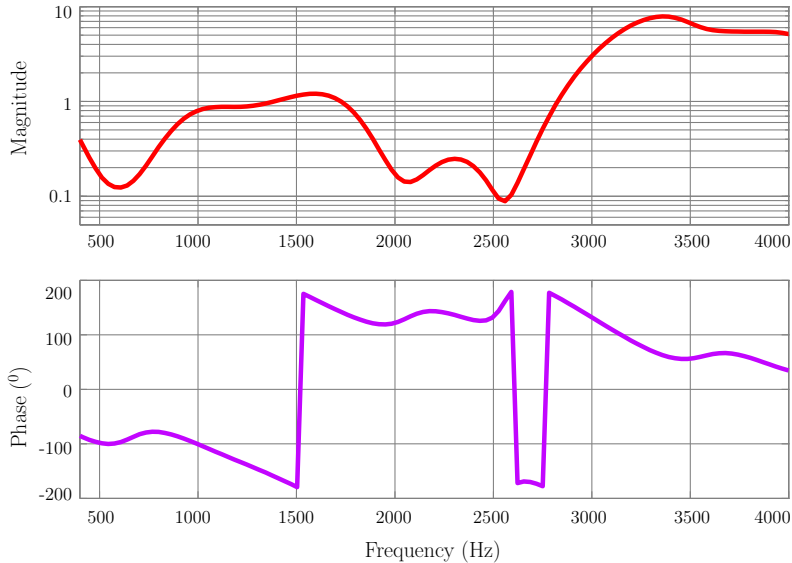}
\caption{Transfer function $\Txhf$ from channel \x\ to channel \h\ used for the simulations.}
\label{fig:TransFns}
\end{figure}

\begin{figure}[tb]
\centering
\includegraphics[width=4.35in]{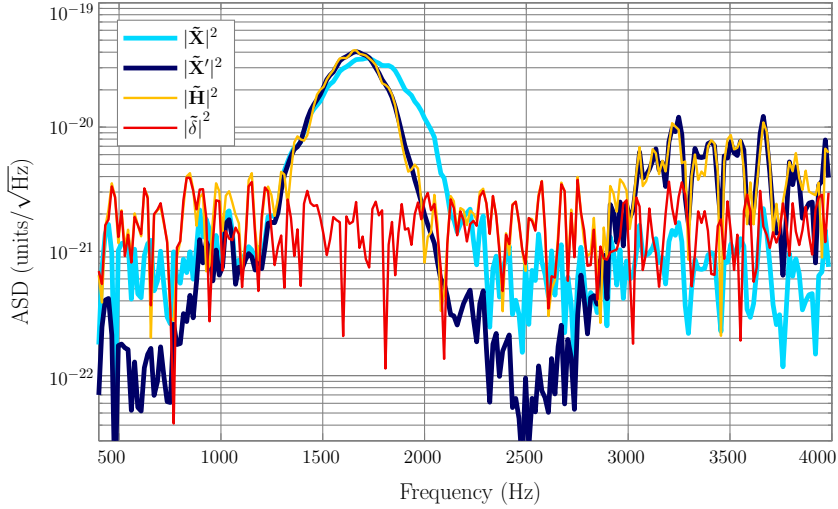}
\caption{Amplitude spectral densities of $\bX$, $\bX'$, $\bH$ and $\btdelta$
in the presence of a sine-Gaussian injection. It can be seen that $\btdelta$ 
contains no trace of the injected signal. The injected sine-Gaussian has a central 
frequency of 1715 Hz and SNR of 140 in channel \x.}
\label{fig:ASDXHdelta}
\end{figure}

Let us define some terminology. The rejection probability is the {\it probability} of a trigger
originating from channel \x\ being vetoed using the method described above. The veto efficiency
is the fraction of such triggers that are {\it actually} vetoed. In order to demonstrate that 
the obtained veto efficiency is in good agreement with the rejection probability, we simulate a 
population of bursts in channel \x\ and \h\ such that these are consistent with the transfer
function from channel \x\ to \h. We then try to veto these triggers after choosing a specific
rejection probability. If all of our assumptions are valid, the fraction of vetoed triggers 
among this population should be equal to the rejection probability. 

We generate a data stream of Gaussian white-noise and `inject' Gaussian-modulated sinusoidal 
waveforms in to it; this forms our channel \x. This data stream is filtered using a time-domain 
filter and some extra noise is added to it. This constitutes our channel \h\ (the `extra' noise
being the component of $\bH$ orthogonal to $\bX'$). The frequency response of the filter is 
the transfer function from \x\ to \h. The transfer function used in this simulation is shown 
in Figure~\ref{fig:TransFns}, which is quite similar to one particular transfer function measured 
in \geo. The injected sine-Gaussians are of the form:
\begin{equation}
s(t) = s_\mathrm{rss}\left(\frac{2 f_0^2}{\pi}\right)^{1/4} \, \sin\left[2\pi f_0(t-t_0)\right] \, 
\exp\left[-(t-t_0)^2/\tau^2\right],
\end{equation}
where $f_0$ is the central frequency of the waveforms and $t_0$ is the time corresponding 
to the peak amplitude. We setup the envelope width as $\tau = 2/f_0$. The corresponding 
quality factor is $Q \equiv \sqrt{2}\pi f_0\tau = 8.9$ and bandwidth is 
$\Delta f = f_0/Q \simeq 0.1 f_0$. The quantity $s_\mathrm{rss}$ is the root-sum-squared (RSS) 
amplitude:
\begin{equation}
\left[\int_{-\infty}^{\infty} s^2(t) \, \dt \right]^{1/2} = s_\mathrm{rss}.
\end{equation}
The $s_\mathrm{rss}$ is varied so that the SNR~\footnote{The SNR, $\rho$, of a burst trigger 
is defined by $\rho^2 \equiv 2 \int \left(|\tilde s(f)|^2/|\tilde n(f)|^2\right) \df$ where 
$\tilde s(f)$ and $\tilde n(f)$ represent the Fourier transforms of the signal and the 
underlying noise, respectively, and the integration is carried out over the bandwidth 
(positive frequencies) of the burst waveform.} 
of the injections ranges from $\simeq$ 6 to $\simeq$ 500 in channel \x, and the central 
frequency is randomly chosen from the interval (432 Hz, 3008Hz). 

As an illustration, the amplitude spectral densities of $\bX$ (data vector in channel \x), 
$\bX'$ (data vector in \x, mapped to \h), $\bH$ (data vector in channel \h) and $\btdelta$ 
(component of $\bH$ that is orthogonal to $\bX'$) in the presence of a particular 
sine-Gaussian injection are shown in Figure~\ref{fig:ASDXHdelta}. 
The injected sine-Gaussian has a central frequency of 1715 Hz and SNR of 140 in channel \x.
As expected, $\btdelta$ contains no trace of the injected signal. 

\begin{figure}[t]
\centering
	\includegraphics[width=5in]{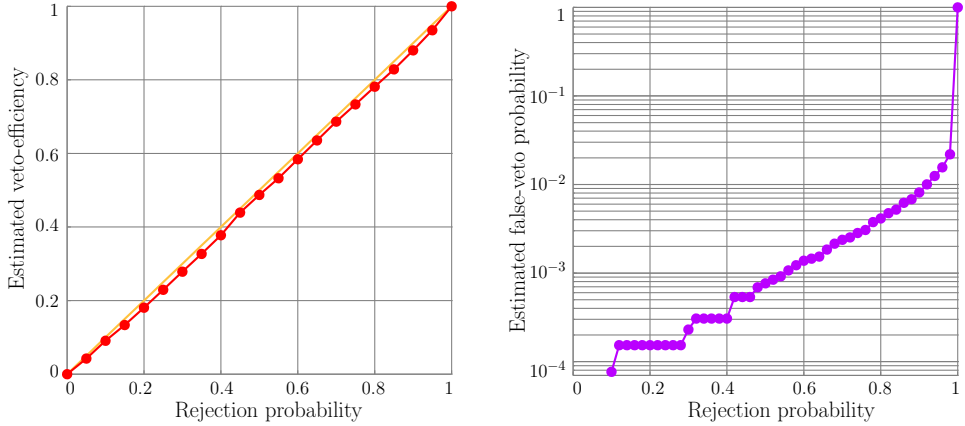}
\caption{[{\it Left}]: Estimated veto efficiency plotted against the rejection probability. 
[{\it Right}]: Estimated false-veto probability plotted against the rejection probability.}
\label{fig:VetoEff}
\end{figure}

The fraction of vetoed events among the injections is plotted against the rejection probability 
corresponding to the chosen threshold in Figure~\ref{fig:VetoEff} (left). It can be seen that 
the obtained veto efficiency is in very good agreement with the chosen rejection probability. 

There is a non-zero probability for an instrumental burst which does not originate in \x, 
or a true GW burst, to be vetoed using this method. But, since the
probability densities of neither GW bursts nor instrumental bursts are known {\it a priori}, 
there is no rigorous method for estimating this `false-rejection'/`false-veto' probability. 
As a plausible estimation, we inject two populations of sine-Gaussian waveforms with random 
parameters into two data streams of white-noise (so that the waveforms in channel \x\ and \h\
are inconsistent with the transfer function from \x\ to \h). We then try to veto these triggers 
using the transfer function from \x\ to \h. The estimated `false-veto fraction' is 
plotted against the chosen rejection probability in Figure~\ref{fig:VetoEff} (right). 
This suggests that veto efficiencies of $\geq$ 92 \% can be achieved with a false-veto 
probability of $\leq$ 1\%. 

\section{An alternative method: `trigger mapping'} 
\label{sec:EvntProjVeto}
Although the above described method is rigorous and makes use of the complete 
information contained in the data, the method can be computationally expensive
because it involves reprocessing of the time series data from the two channels. 
It may be noted that the whole data stream is processed by the ETG in the first
place and a condensed form of the information about the burst waveform is stored
(which is often known as `metadata'). In this section, we develop a strategy to 
veto the spurious triggers in channel \h\ by comparing them with the set of triggers
in channel \x, making use of the transfer function from \x\ to \h\ as well as the 
information extracted by the ETG. Although this method is not as rigorous as the 
previous one, the advantage is that this does not require the reprocessing of 
time-series data and hence is computationally inexpensive.  

Let $\{\Ex\}$ and $\{\Eh\}$ denote the set of burst triggers in channel \x\ and channel 
\h, respectively. Let us assume that each event, $\bf E$, is parametrized by its central 
frequency $E^f$, amplitude $E^a$ and time-of-occurrence $E^t$. It is useful to think of 
$\bf E$ as a point in a three-dimensional parameter space with coordinates $(E^a,E^f,E^t)$. 
Using the transfer function from \x\ to \h, we can predict how a certain event $\Ex$ 
would appear in \h. In other words, we map the event $\Ex$ to the space of $\Eh$, making
use of the transfer function from \x\ to \h. In order to veto an event $\Eh$ in channel 
\h, we check whether any of the `mapped' $\Ex$ triggers are consistent with $\Eh$ in 
time-of-occurrence, central frequency and amplitude. Indeed, the precise definitions of 
these parameters depend upon the ETG, and hence we make use of these definitions in order
to map the burst triggers from one channel to the other. 

\subsection{Mapping the burst triggers}

\begin{figure}[t]
\centering
\includegraphics[width=2.in]{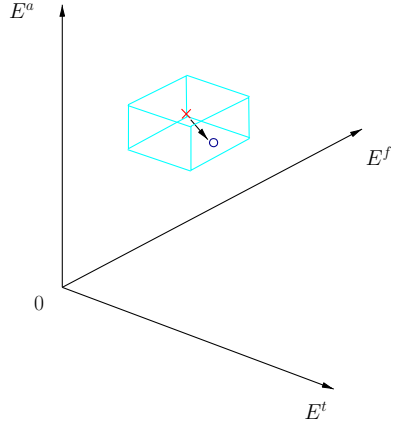}
\caption{A schematic diagram of the parameter space of burst triggers. The cross 
represents a trigger in channel \h\ and the circle represents a trigger in channel \x, 
mapped to channel \h. A distance vector ${\bf w}$ is drawn from one point to another. 
The consistency volume chosen to veto the trigger is also shown. }
\label{fig:ParamsSpace}
\end{figure}

The burst detection algorithm called HACR \cite{Siong,Bala} is used to generate the 
burst triggers. HACR makes a time-frequency map of the data and identifies time-frequency 
pixels containing excess power which are statistically unlikely to be associated with the 
underlying noise. HACR then proceeds to cluster the neighboring pixels containing excess 
power to form an `event'. The central frequency and time-of-occurrence of the burst-event 
are estimated by a weighted averaging of the pixel coordinates. This is equivalent to the 
calculation of the center-of-mass of an extended object where the signal power in a pixel 
serves as the `mass' term. HACR also estimates the total power contained in all the pixels 
belonging to a particular event, and the peak-power of the event. The square root of the 
total power is taken as the characteristic amplitude $\Exa$ of the event. 
 
The ETG does not reproduce the complete physical properties of a burst waveform. Instead, 
the ETG tries to parametrize the underlying waveform using a set of quantities like the 
central frequency, amplitude, bandwidth, duration, etc. Considering the fact that we are 
mostly looking at short-lived, band-limited bursts, we approximate the power spectrum of 
the underlying burst waveform in channel \x\ to a Gaussian function. For example, the power 
spectrum of the waveform associated with a trigger ${\bf E}_{\rm X}$ is approximated to a 
Gaussian function $G(f)$ such that 
\begin{equation}
\int_{f_1}^{f_2} G(f) \, \df = \big( E^a_{\rm X} \big)^2,
\label{eq:gauss}
\end{equation}
where the limits of integration are defined by bandwidth $\Exb$ of the burst, i.e. 
$f_1 = \Exf - \Exb/2$ and $f_2 = \Exf + \Exb/2$. Since the peak-power of the 
burst is also estimated by the ETG, the `spread' of the Gaussian function can 
be calculated by solving Eq.(\ref{eq:gauss}). The power spectrum is `deformed' 
by the transfer function $\Txhf$ when the glitch makes its way to channel \h; this 
we denote by
\begin{equation}
\hat G(f) =  G(f) \,|\Txhf|^2.
\end{equation}
Given the transfer function $\Txhf$, the burst triggers in channel \x
can be mapped to channel \h\ in the following way:
\begin{eqnarray}
\Exa {^\prime} = \left[ \int_{f_1}^{f_2}\hat G(f)\,\df \right]^{1/2} \,, 
\label{eq:EvntProjA} \\
\Ext {^\prime} = E^t_{\rm X} \, +  \frac{\int_{f_1}^{f_2} \hat G(f)\,\lambda(f)\, \df}
{\int_{f_1}^{f_2} \hat G(f)\,\df} \,,
\label{eq:EvntProjF} \\
\Exf {^\prime} = \frac{\int_{f_1}^{f_2} \hat G(f)\,f\,\df} 
 {\int_{f_1}^{f_2} \hat G(f)\,\df}\,.
\label{eq:EvntProjF}
\end{eqnarray}
In the above expression, $\lambda(f)$ is the frequency-normalized phase-delay (time-lag) 
of the transfer function $\Txhf$. i.e, 
\begin{equation}
\lambda(f) = \frac{1}{2\pi f} \, \phi \big(\Txhf\big)\,.
\end{equation}
where $\phi(.)$ denotes the phase of a complex quantity.

\subsection{Identifying consistent events}

\begin{figure}[t]
\centering
\includegraphics[width=5.1in]{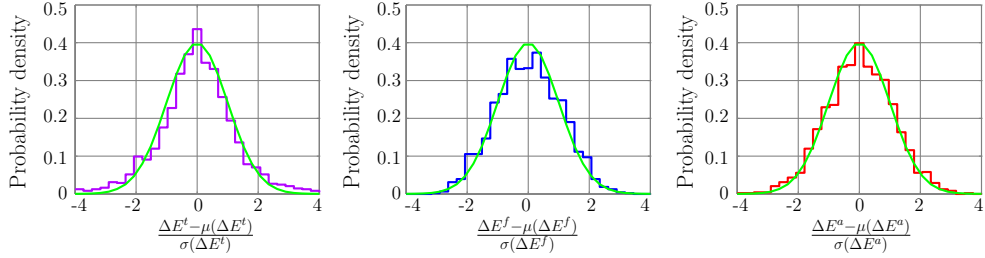}
\caption{Distribution of HACR errors (after subtracting the mean and normalizing by the 
standard deviation) in estimating the parameters of the injected sine-Gaussian waveforms. 
Also plotted is the probability density of Normal distribution with mean 0 and variance 1.}
\label{fig:HACRerror}
\end{figure}

\begin{figure}[t]
\centering
\includegraphics[width=5.1in]{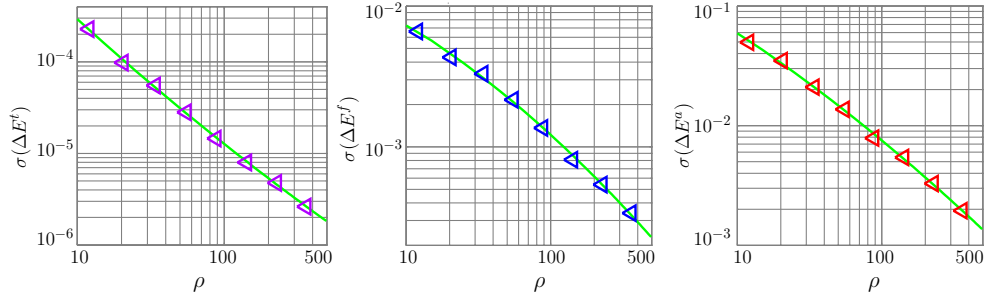}
\caption{Standard deviation of the distribution of the HACR errors (in estimating the
parameters of the injected sine-Gaussian waveforms) plotted as a function
of the SNR of the triggers. Also shown are the polynomial fits to the data.}
\label{fig:ErrorSigmas}
\end{figure}

In order to veto a trigger, $\Eh$, in channel \h, we check whether any of the `mapped' triggers 
($\Ex'$) from channel \x, are `sufficiently close' to it. To be explicit, we define a 
vector, ${\bf w}$, connecting the two points in the parameter space of \h\ triggers, 
\begin{equation}
{\bf w} \equiv  \Eh - \Ex',
\end{equation} 
and require that it has a sufficiently small `length'. This length is assigned to the
vector by calculating the fractional volume enclosed by a three-dimensional Gaussian 
envelope of width $\sigma({\bf w})$. This is explained below. 

Let $\dE$ denote the errors in the ETG in the estimation of parameters associated with the 
event $\bf E$. In the absence of any systematic biases, the errors $\dE$ can be assumed to be 
drawn from multivariate Normal distributions of zero mean and standard deviation $\sigma(\dE)$,
where the standard deviation is an exponentially decreasing function of the SNR. $\sigma(\dE)$ 
can be estimated by injecting known waveforms in to the data and comparing the trigger-parameters 
estimated by the ETG to the actual parameters of the injected waveforms.

Once the errors $\dEx$ associated with $\Ex$ are estimated, they can be mapped to 
the space of \h\ triggers using Eq.(\ref{eq:EvntProjA}-\ref{eq:EvntProjF}) 
by a linear approximation of the error propagation~\cite{BevingtonRobinson}. 
Then the components of ${\bf w}$ will be distributed according to Normal distributions
of zero mean and the following variance:
\begin{eqnarray}
\sigma^2({\bf w}) &=& \sigma^2(\dEh) + \sigma^2 \big(\dEx') , 
\end{eqnarray}
where $\dEx'$ denotes the errors $\dEx$, mapped to the space of \h\ triggers. 
Thus the joint probability density of the vector ${\bf w}$ comprising of the three random
variables $(w^a,w^f,w^t)$ is given by the three-dimensional Gaussian function
\begin{equation}
f({\bf w}) = \frac{1}{(2 \pi)^{3/2} \, \sigma(w^f) \, \sigma(w^a) \, \sigma(w^t)}\exp 
\left(-\frac{1}{2}{\bf w}^{\rm T}\, {\sf C}^{-1} \,  {\bf w} \right),
\end{equation}
where ${\bf w}^{\rm T}$ denotes the transpose of $\bf w$. Assuming that the errors are 
uncorrelated, we write the covariance matrix $\sf C$ as,
\begin{equation}
\sf C = \left[ 
\begin{array}{ccc} 
\sigma^2(w^f) & 0 & 0 \\
0 & \sigma^2(w^a) & 0 \\
0 & 0 & \sigma^2(w^t) 
\end{array} \right] \,.
\end{equation}
This enables us to set a threshold for ${\bf w}$ for vetoing an event ${\bf E}_{\rm H}$. 
In order to veto an event, we require that
\begin{equation}
|{\bf w}| \leq {\bm \tau}.
\end{equation}
The components ($\tau^a,\tau^t,\tau^f$) of the `threshold vector' $\bm \tau$ 
are related to the rejection probability $\psi$ by
\begin{equation}
\psi = 8 \int_{0}^{\tau^f}\int_{0}^{\tau^a}\int_{0}^{\tau^t} \,
f({\bf w}) \, \dwf \, \dwa \, \dwt.
\end{equation}

It can be seen that, by choosing a particular threshold vector $\bm \tau$, we are 
defining a consistency volume around each burst trigger in channel \h. If one of the
mapped triggers from channel \x\ falls in to this volume, we veto the \h\ trigger. This 
is schematically illustrated in Figure~\ref{fig:ParamsSpace}.   

\subsection {Software injections}

\begin{figure}[t]
\centering
\includegraphics[width=4.5in]{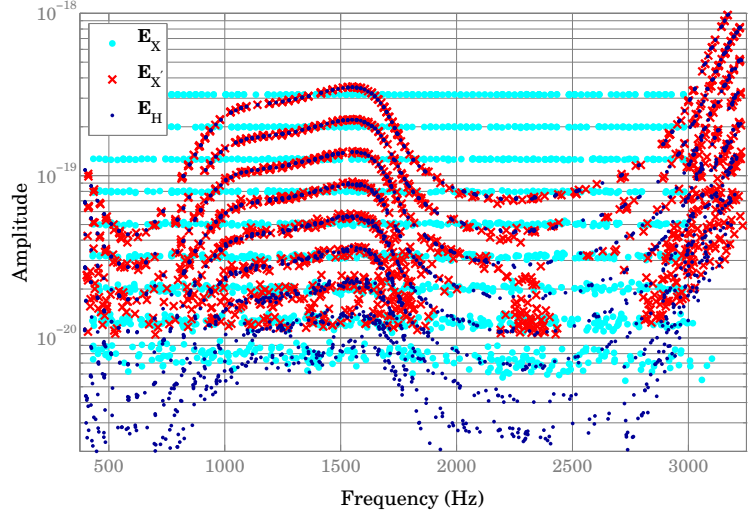}
\caption{The RSS amplitude, $E^a$, of the burst-events plotted against the central 
frequency, $E^f$. Shown in the Figure are three population of events: events in channel 
\x\ ($\Ex$); channel \x\ events mapped to channel \h\ ($\Ex'$); and events in channel 
\h\ ($\Eh$).} 
\label{fig:InjFAplot}
\end{figure}

In order to estimate the variance of the distribution of HACR errors in estimating the 
parameters of the waveform, sine-Gaussian waveforms are injected into white noise and the 
HACR ETG is run over the data. The parameters estimated by the ETG are compared to the 
injected parameters in order to calculate the errors in the parameter-estimation. Distributions 
of the errors (after subtracting the mean and normalizing by the standard deviation) are 
plotted in Figure~\ref{fig:HACRerror}, along with the probability density of the Normal
distribution with mean 0 and variance 1. The standard deviation of the errors in the 
estimation of different parameters are plotted as a function of the SNR of the triggers 
in Figure~\ref{fig:ErrorSigmas}. It can be seen that, to a very good approximation, the 
standard deviation is an exponentially decreasing function of the SNR. The eight data 
points in the plots in Figure~\ref{fig:ErrorSigmas} correspond to eight different values 
of the RSS amplitude used for the injections. We fit the data with a simple polynomial 
fit and take it as the functional form of the standard deviation. 

We simulate two populations of bursts in channel \x\  and \h, as described in Section~
\ref{sec:SWinj}. The two data streams are processed by HACR and two sets of triggers 
$\{{\bf E}_{\rm X}\}$ and $\{{\bf E}_{\rm H}\}$ are generated. The \x\ triggers are
mapped to channel \h\ using the transfer function from \x\ to \h. Figure~\ref{fig:InjFAplot} 
shows the characteristic amplitude $E^a$ of the three population of triggers ($\Ex,\Ex'$ 
and $\Eh$) plotted against the central frequency $E^f$. The injected events in channel
$X$ span 9 different amplitudes. The $\Ex'$ triggers and $\Eh$ triggers can be seen to 
fall nicely in to the shape of the transfer function.  

\begin{figure}[t]
\centering
\includegraphics[width=5in]{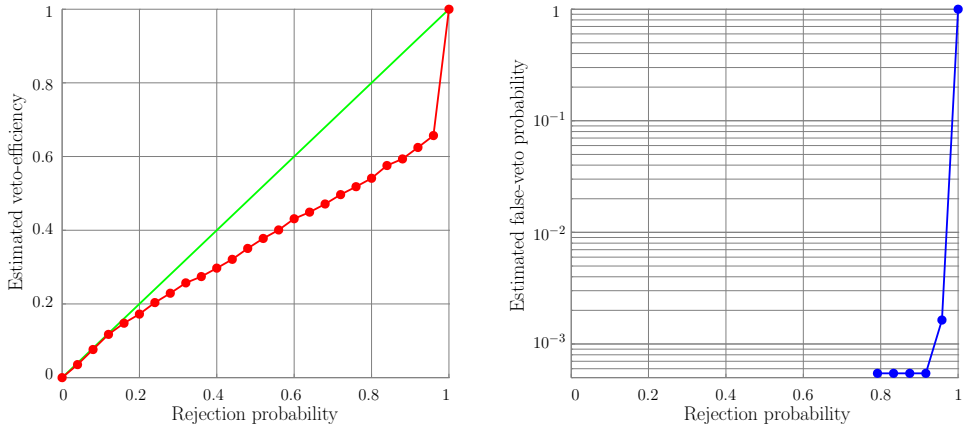}
\caption{[{\it Left}]: Estimated veto efficiency plotted against the rejection probability. 
[{\it Right}]: Estimated false-veto probability plotted against the rejection probability.}
\label{fig:EPVetoEff}
\end{figure}

The veto analysis is repeated with different thresholds. The estimated veto efficiency 
is plotted against the rejection probability corresponding to the chosen thresholds in 
Figure~\ref{fig:EPVetoEff}. Although the estimated rejection efficiency roughly agrees 
with the predicted rejection probability, the effect of relying on a number of assumptions 
can be immediately seen. The main source of error in the analysis comes from the inaccurate
parametrization of the errors in ETG in the parameter-estimation of the burst waveforms. 
This can be different for different ETGs.  

As described in Section \ref{sec:SWinj}, we also estimate the false-veto probability by 
injecting sine-Gaussian waveforms with random parameters in to the two data streams and 
preforming the veto analysis. The estimated false-veto probability is plotted in Figure~
\ref{fig:EPVetoEff} (right) as a function of the rejection probability. This exercise
suggests that a veto efficiency of $\sim 70\%$ can be achieved at the cost of a false-veto 
probability $\sim 1\%$.

\section{Summary and outlook}

In this paper we have formulated and demonstrated a strategy to veto spurious noise 
transients that appear in the output of interferometric GW detectors. This method is novel
in the sense that it makes use of the measured coupling between an instrumental sub-system 
and the detector output in order to veto a spurious trigger. The basic idea is that the
noise recorded in an instrumental channel \x\ can be mapped in to the detector output 
(channel \h) using the transfer function from \x\ to \h. This enables us to project the 
noise at the detector output in to two orthogonal directions -- along, and orthogonal 
to, the mapped noise from channel \x. If a non-stationarity in \h\ is originated from 
channel \x, it leaves the statistics of the component of \h\ orthogonal to the `mapped' \x unchanged. 
This can be verified by a statistical hypothesis testing. We have also proposed a less-rigorous, 
but computationally inexpensive alternative to the above method. In this method, the parameters 
of the triggers in channel \x\ are compared to the parameters of the triggers in
channel \h\ to see whether a trigger in channel \x\ can explain a trigger in channel \h.  

Work is ongoing to incorporate the veto in the online data characterization pipeline 
of the \geo\  detector~\cite{GeoReports}. So far, we have not explored the effect of the errors 
in measuring the transfer function. This certainly needs to be considered when the veto is 
applied to the actual data characterization/analysis pipeline. We leave that as future work. 
Also, the assumption that the transfer function is time-invariant is also not strictly 
true. Transfer functions in actual detectors can vary in time. This issue can be addressed 
by making repeated measurements of the transfer function, and tracking its evolution by 
continuously injecting and measuring spectral lines at certain frequencies. This is 
described in~\cite{NoiseProj} and is already being practiced in \geo. It may also be noted 
that the noise in the present-generation interferometers are not stationary Gaussian, 
and exhibits tails in the distribution. An actual implementation of this method should 
take this also in to account. 

The `trigger mapping' veto needs to make certain assumptions about 
the power spectrum of the glitch in channel \x. The assumption that we made in the paper, 
that the power spectrum can be approximated by a Gaussian function, should be verified 
against real-life glitches. It might also be possible to make assumptions which are closer 
to the reality, using better parametrization of the underlying waveforms.  

The results of the two veto methods presented in this paper may not be compared directly.
The very choice of sine-Gaussian waveforms for the software injections especially favour the 
trigger mapping method, as they satisfy the assumption about the power spectrum of the
burst waveforms that we used in the analysis. Yet, the results show that the `noise projection'
method performs better. It may be noted that the trigger mapping method also depends on the 
accuracy of the parameter-estimation of the ETG while the `noise projection' veto relies 
only on the accuracy with which the transfer function is known.

\section*{Acknowledgments}

The authors would like to thank all the members of \geo\ group of the Albert Einstein 
Institute for very useful discussions, and Benno Willke for his comments on the manuscript. 
P.A. also acknowledges enlightening discussions with Jan Harms, Reinhard Prix and 
B. S. Sathyaprakash.  

\bigskip


\end {document}